\documentclass[twoside,twocolumn,english,aps,showpacs,prl,superscriptaddress]{revtex4-1}
\usepackage[T1]{fontenc}
\usepackage[utf8]{inputenc}
\usepackage{color}
\usepackage{bm}
\usepackage{amsmath}
\usepackage{amsthm}
\usepackage{amssymb}
\usepackage{graphicx}
\usepackage{esint}
\usepackage{mathrsfs}
\usepackage{booktabs}
\usepackage{multirow}
\usepackage{makecell}
\usepackage{placeins}
\usepackage{diagbox}
\usepackage{tabularx}

\usepackage[plainpages = false, pdfpagelabels, 
                 bookmarks,
                 bookmarksopen = true,
                 bookmarksnumbered = true,
                 breaklinks = true,
                 linktocpage,
                 colorlinks = true,
                 linkcolor = blue,
                 urlcolor  = blue,
                 citecolor = blue,
                 anchorcolor = green,
                 hyperindex = true,
                 hyperfigures
                 ]{hyperref} 
\usepackage{dcolumn}   % needed for some tables
\usepackage{array, makecell}

\begin{document}
\title{Mitigating crosstalk and residual coupling errors in superconducting quantum processors using many-body localization}

\author{Peng Qian}
\affiliation{Beijing Academy of Quantum Information Sciences, Beijing 100193, China}
\affiliation{State Key Laboratory of Low Dimensional Quantum Physics, Department of Physics, Tsinghua University, Beijing, 100084, China}

\author{Hong-Ze Xu}
\affiliation{Beijing Academy of Quantum Information Sciences, Beijing 100193, China}
\affiliation{State Key Laboratory of Low Dimensional Quantum Physics, Department of Physics, Tsinghua University, Beijing, 100084, China}

\author{Peng Zhao}
\affiliation{Beijing Academy of Quantum Information Sciences, Beijing 100193, China}

\author{Xiao Li}
\affiliation{Department of Physics, City University of Hong Kong, Kowloon, Hong Kong SAR, China}

\author{Dong E. Liu}
\email{Corresponding to: dongeliu@mail.tsinghua.edu.cn}
\affiliation{State Key Laboratory of Low Dimensional Quantum Physics, Department of Physics, Tsinghua University, Beijing, 100084, China}
\affiliation{Beijing Academy of Quantum Information Sciences, Beijing 100193, China}
\affiliation{Frontier Science Center for Quantum Information, Beijing 100184, China}
\affiliation{Hefei National Laboratory, Hefei 230088, China}
%\pacs{ }
\begin{abstract}
Addressing the paramount need for precise calibration in superconducting quantum qubits, especially in frequency control, this study introduces a novel calibration scheme harnessing the principles of many-body localization (MBL). While existing strategies, such as Google's snake algorithm, have targeted optimization of qubit frequency parameters, our MBL-based methodology emerges as a stalwart against noise, notably crosstalk and residual coupling errors, thereby significantly enhancing quantum processor fidelity and stability without necessitating extensive optimization computation. Not only does this approach provide a marked improvement in performance, particularly where specific residue couplings are present, but it also presents a more resource-efficient and cost-effective calibration process. The research delineated herein affords fresh insights into advanced calibration strategies and propels forward the domain of superconducting quantum computation by offering a robust framework for future explorations in minimizing error and optimizing qubit performance.
\end{abstract}
\maketitle

%Superconducting quantum qubits have shown immense potential for the development of quantum computers due to their high scalability and compatibility with existing semiconductor technologies. However, their performance is heavily dependent on precise calibration, particularly in terms of qubit frequency control. Recently, Google used the intriguing snake algorithm to optimize the frequency parameters of superconducting qubits, achieving good performance. In this study, we propose a new scheme for parameter selection based on the physical principles of many-body localization (MBL). This approach has the potential to provide convenience and cost savings in the calibration process while demonstrating good performance without requiring a large amount of optimization computation.

\textbf{\em Introduction--} The calibration \cite{Arute2019Supremacy,snake,DAG,PhysRevLett.127.180501,majumder2020real,xu2023digital} of superconducting quantum computers is a crucial step to ensure the device's precise calculations. As the scope of applications expands, the calibration accuracy and efficiency requirements also increase. The exponential growth in the number of qubits also results in an increase in the space of system control parameters, making it more challenging to find a fast and effective calibration solution.

In recent years, progress has been made in superconducting quantum computing calibration. For instance, Google uses the feature of neighbor coupling to give the snake algorithm~\cite{snake,Arute2019Supremacy,DAG}, while IBM has developed a data-based and self-learning method to calibrate the qubits, automatically adjusting the parameters of superconducting quantum computers to optimize performance \cite{PhysRevApplied.15.034080}. However, there are still challenges, such as calibration often requiring a lot of time and computing resources, especially with an increase in the number of qubits and gate operands. This not only increases the cost of calibration but also affects the feasibility of practical applications. Furthermore, superconducting chip calibration typically needs to be performed in a low temperature and low noise environment, yet environmental interference, such as thermal fluctuations, vibrations, and electromagnetic radiation, can lead to calibration errors and instability \cite{krantz2019quantum,circuitqed}. Thus, finding more robust parameters to mitigate these instabilities remains a difficult problem.

Frequency parameter control \cite{Arute2019Supremacy,snake,PhysRevLett.127.180501,majumder2020real,xu2023digital,wittler2021integrated,motzoi2009simple,krantz2019quantum,Yan2018coupler} of superconducting qubits is one of the most crucial aspects that require attention. The frequency of a qubit is responsible for determining its lifetime and coherence, as well as the accuracy of individual qubit gates and the coupling between adjacent qubits. Moreover, since there is a range of classical or quantum crosstalk between the qubits, the selection of qubit frequency is crucial in this regard. The optimization of qubit frequency selection can help to suppress crosstalk, thereby enhancing the performance of superconducting qubits in terms of fidelity, stability, and other related aspects\cite{Arute2019Supremacy,snake,DAG,PhysRevLett.127.180501,majumder2020real,xu2023digital,wittler2021integrated,motzoi2009simple,PhysRevApplied.12.064022,PRXQuantum.2.040313,PhysRevApplied.20.024070,braumuller2022probing,dai2022optimizing,karamlou2023probing}.

Superconducting quantum chips, being many-body systems, inherently exhibit characteristics of many-body physics \cite{berke2022transmon}. Among these, many-body localization (MBL) has drawn significant interest in recent years \cite{basko2006metal,nandkishore2015many,ALET2018498,Abanin201700169,RevModPhys.91.021001}. 
%\textcolor{red}{(* REMOVE MBL is a quantum phenomenon in which certain systems resist thermalization, remaining localized and maintaining a coherent state, even under weak coupling, as opposed to progressing towards a thermal state. END*)}
MBL is a novel dynamical phase in which a disordered isolated quantum many-body system may remain localized even in the presence of finite interactions. 
This contrasts with thermalization, where a system eventually transitions into a uniform mixed state devoid of coherence. In quantum computing, preserving the calculated state until its retrieval, or mitigating undesired coupling between qubits during gate operations, is paramount. As such, the MBL effect could potentially enhance the energy efficiency of quantum chips \cite{berke2022transmon}.

In this work, we advocate for the utilization of the Many-Body Localization (MBL) parameter interval as the initial calibration value by tuning the frequencies of superconducting qubits. This methodology, pivotal in improving the fidelity of quantum processors, acts as a stalwart against noise—predominantly significant crosstalk and residual coupling errors—thereby realizing a marked enhancement in performance. Numerical simulations further elucidate the merits of employing MBL-centric calibration for superconducting qubits, especially those with specific residue couplings, compared with non-MBL-based calibration strategies. Our research unveils nuanced insights into the calibration process, critically relevant to the domain of quantum computing.

%In this work, we propose the use of the Many-Body Localization (MBL) parameter interval, that is to control the frequencies for superconducting qubits, as the initial calibration value. This methodology can significantly enhance the fidelity of the quantum processors to battle noise,  especially crosstalk and residual coupling errors, yielding a substantial improvement in performance. We also present numerical simulation results that underscore the advantages of employing MBL-based calibration for superconducting qubits with certain coupling residues, in contrast to non-MBL-based calibration. Our research offers insights into the calibration process integral to quantum computing.

\begin{figure}[htbp]
\centering
\includegraphics[keepaspectratio=true,width=\columnwidth]{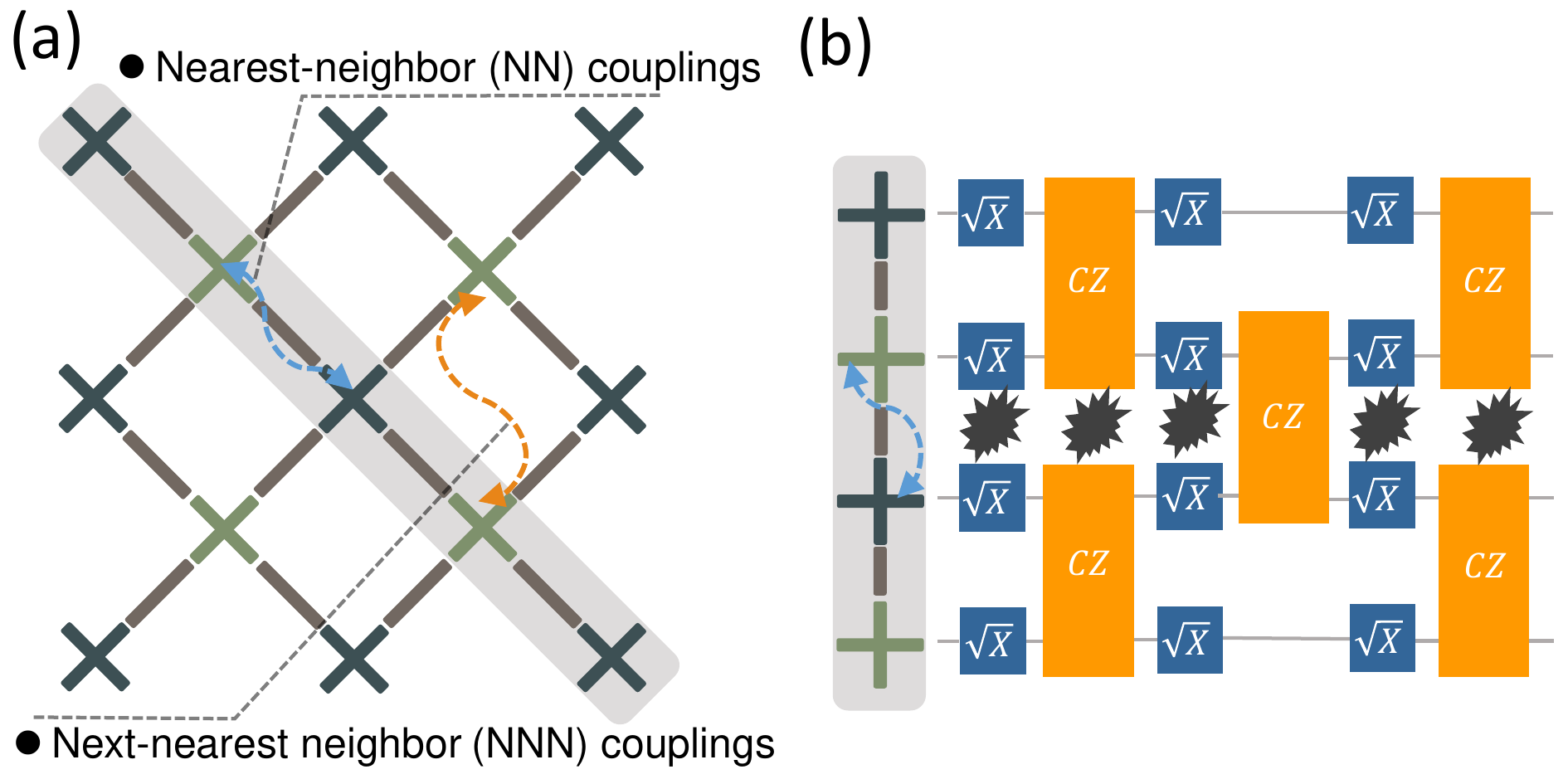}
\caption{Residual couplings in a typical two dimensional qubit lattice. (a) The cross-shaped vertices (green) denote qubits, and the strip-shaped edges (grey) between adjacent qubits represent tunable couplers. Generally, both the residual NN (blue dashed lines with arrows) and NNN (orange dashed lines with arrows) couplings can be suppressed in qubit lattices  featuring tunable coupling. (b) Besides the desired couplings, residual NN or NNN couplings, which are caused by qubit or coupler frequency miscalibrations, can exist. This will lead to crosstalk effect, which can inadvertently influence both the single-qubit and two-qubit operations, and thus pose significant challenges to the integrity and accuracy of quantum computations. } \label{fig: residual_coupling}
\end{figure}

\textbf{\em Superconducting quantum processors and residual couplings--}. For illustration purposes only, hereafter, we focus on superconducting quantum processors based on transmon qubits~\cite{Koch2007transmon}. For such processors, transmon qubits can be modeled as anharmonicity oscillators and the interqubit couplings can be adjusted by using tunable couplers~\cite{Yan2018coupler}. Accordingly, the processor can be described by the following Bose-Hubbard Hamiltonian~\cite{krantz2019quantum,PhysRevB.87.134202}
\begin{equation}
\begin{aligned}\label{eq:systemH}
&H=H_{0}+H_{int},
\\&H_{0}=\sum_{j}\big[\omega_{j}a_{j}^{\dagger}a_{j}
+\frac{V_{j}}{2}a_{j}^{\dagger}a_{j}(a_{j}^{\dagger}a_{j}-1)\big],
\\&H_{int}=\sum_{j,k}h_{jk}(a_{j}^{\dagger}a_{k}+a_{j}a_{k}^{\dagger}),
\end{aligned} 
%\label{fund}
\end{equation}
where $\omega_{j}$ and $V_{j}$ denote the qubit frequency and anharmonicity, respectively, $h_{jk}$ is the interqubit coupling strength, and $a_{j}\,(a_{j}^{\dagger})$ represent the annihilation (creation) operator. 

Generally, in the above quantum processor, two-qubit gates can be realized by tuning qubits into a specific resonance condition (e.g., when two qubits are put on resonance, a two-qubit iSWAP-type gate can be achieved) and by turning on the interqubit couplings~\cite{Foxen2020Gate}, while single-qubit gates are activated by adding microwave drives, giving rise to the following drive Hamiltonian
\begin{equation}
\begin{aligned}\label{eq:driveH}
H_{d,j}=\epsilon_{j}(t)(a_{j}^{\dagger}+a_{j}),
\end{aligned}
\end{equation}
where $\epsilon_{j}(t)$ is given as
\begin{equation}
\begin{aligned}\label{eq:drive_pulse}
\epsilon_{j}(t)=\epsilon^{x}(t)\cos(\omega_{d}t+\phi_{0})+\epsilon^{y}(t)\sin(\omega_{d}t+\phi_{0}).
\end{aligned}
\end{equation}
Considering the weak transmon anharmonicity, here, the derivative removal by adiabatic gate (DRAG) pulse~\cite{Motzoi2009Drag} is used for suppressing leakage outside the qubit computational subspace. The choice of the phase $\phi_{0}$ can be used to generate the specific gate operations, e.g. a phase of $ \phi_{0} = 0 $ corresponds to an $ X $ gate, while $ \phi_{0} = \frac{\pi}{2} $ is associated with a $ Y $ gate. 
In practical implementations, additional single-qubit gates such as $ \sqrt{X} $, $ \sqrt{Y} $, and $ \sqrt{W} $, where $ W = \frac{X + Y}{\sqrt{2}} $, can be also executed~\cite{krantz2019quantum,Arute2019Supremacy}. 

For certain implementations of multi-qubit operations in superconducting systems, the drive frequency $ \omega_{d} $ is tuned to match or come close to the resonant frequency of the target qubit. In the realm of superconducting quantum computation, the Controlled-Z (CZ) and iSWAP gates are commonly used two-qubit operations. 
Nonetheless, when one takes into account the practicalities of physical implementation, especially in scenarios with complex Hamiltonians, such as the one represented by Equation~(\ref{eq:systemH}), that involve elevated energy states, the fsim gate emerges as a more pragmatic choice~\cite{krantz2019quantum,Arute2019Supremacy,kivlichan2018quantum}. 

In the execution of two-qubit gates within superconducting quantum systems, it is standard procedure to engage the coupler connecting the pair of qubits, as shown in Fig.~\ref{fig: residual_coupling}(a). This allows for their evolution in accordance with a predefined calibration protocol. Once this phase is complete, the coupler is disengaged to finalize the gate operation. However, in practical settings, achieving total isolation with the coupler remains a challenge. Consequently, this lack of perfect isolation can inadvertently influence not only the qubits involved during single-qubit operations or while they're idle but also neighboring qubits that aren't directly participating in the specific two-qubit gate operation, as shown in Fig.~\ref{fig: residual_coupling}(b). It's imperative to note that such crosstalk effects, resulting from the coupler's incomplete isolation, can pose significant challenges to the integrity and accuracy of quantum computations.

One of the salient advantages of qubit architectures featuring tunable coupling mechanisms is the capacity to significantly mitigate residual interactions, such as nearest-neighbor (NN) and next-nearest neighbor (NNN) parasitic couplings~\cite{Arute2019Supremacy}, as shown in Fig.~\ref{fig: residual_coupling}(a). However, the comprehensive suppression of these unwanted couplings presents a formidable challenge, particularly in the context of large-scale quantum processors. This complexity arises primarily because the on/off states of interqubit couplings are highly sensitive to variations in the qubit and coupler frequencies ~\cite{google2023suppressing,PRXQuantum.3.020301}. As such, the suppression of extraneous interqubit interactions necessitates precise calibration of both individual qubit frequencies and inter-qubit coupling strengths~\cite{Arute2019Supremacy}. The presence of these residual couplings introduces coherent and correlated errors during quantum computations, posing a significant risk of computational failure.

\textbf{\em Superconducting qubit control and MBL--} 
Many-body localization (MBL) represents a unique non-equilibrium state that challenges conventional understandings provided by equilibrium statistical mechanics and notably breaches the eigenstate thermalization hypothesis (ETH) \cite{Deutsch_2018,PhysRevB.82.174411,ALET2018498}. Unlike typical systems that approach thermal equilibrium at non-zero temperatures, MBL systems resist this trend, retaining their insulating character. This MBL behavior has been explored in various contexts, from random disordered systems \cite{gornyi2005interacting,PhysRevB.77.064426,PhysRevB.97.155133,PhysRevB.102.014310} and quasiperiodic structures \cite{PhysRevB.87.134202,PhysRevB.96.075146,PhysRevLett.121.206601} to Stark potential configurations \cite{PhysRevB.102.054206,PhysRevLett.127.240502,morong_observation_2021}. The properties of MBL systems have been extensively studied via theoretical analysis \cite{basko2006metal,PhysRevLett.122.040601,PhysRevX.5.031032}, numerical simulations \cite{PhysRevB.102.014310,SciPostPhys.5.5.045}, and diverse experimental platforms \cite{bordia2017peridocially, schreiber2015observation,rispoli2019quantum}.

Distinct properties underscore the MBL phase. For instance, instead of adhering to the Wigner-Dyson distribution like thermalized systems, the level spacing in MBL systems aligns more with the Poisson distribution \cite{PhysRevB.93.174202}. Furthermore, MBL states conform to the area law, with entanglement entropy displaying a logarithmic time growth \cite{PhysRevLett.109.017202,PhysRevB.93.060201}. A pivotal attribute is the prolonged retention of initial state information by the MBL phase \cite{PhysRevLett.120.050507,science.aaa7432}, offering potential advantages for quantum storage. Critically, the prethermal regime inherent to MBL systems can slow down undesired quantum correction propagations, presenting a promising avenue to address calibration challenges in superconducting quantum processors.

In this work, we explore the application of MBL phase properties to modulate superconducting quantum processors. Our aim is to mitigate the effects of residual couplings and enhance the preservation of the system's initial state.
In a tunable superconducting qubit system, one can represent the dynamics using the Bose-Hubbard Hamiltonian, 
\begin{equation}
H=\sum \omega_i \hat{n}_i-\sum_i V_i \hat{n}_i\left(\hat{n}_i-1\right)-\sum_{i j} h_{i j}(a_i^{\dagger} a_j+a_i a_j^{\dagger}),
\label{BHH}
\end{equation}
which has been rigorously investigated within the many-body localization (MBL) community \cite{alet2018many,lev2014dynamics,PhysRevB.87.134202,PhysRevLett.122.040606}.
For the convenience of discussion, we fix the parameters $V_i$ and $h_{ij}$. In order to make the system in the MBL phase, the parameter $\omega_i$ can be adjusted in several different ways~\cite{RevModPhys.80.1355,PhysRevB.91.081103,PhysRevA.101.063617,PhysRevLett.125.155701,PhysRevB.105.054203,PhysRevA.101.063617,PhysRevB.82.174411,PhysRevB.77.064426,aubry1980analyticity,PhysRevB.87.134202,wahl_signatures_2019,PhysRevResearch.1.033183,PhysRevB.87.134202,PhysRevB.96.075146,PhysRevLett.121.206601,PhysRevLett.122.040606,RevModPhys.34.645,PhysRevLett.122.040606, PhysRevB.102.054206}.

The realization of MBL phase can be achieved by incorporating both random disorder potentials~\cite{RevModPhys.80.1355,PhysRevB.91.081103,PhysRevA.101.063617,PhysRevLett.125.155701,wahl_signatures_2019,PhysRevB.105.054203,PhysRevA.101.063617,PhysRevB.82.174411,PhysRevB.77.064426} and Stark potentials~\cite{RevModPhys.34.645} into the model. In this work, our emphasis is on an alternative realization mechanism utilizing a quasi-periodic potential~\cite{aubry1980analyticity,PhysRevB.87.134202}, which is characterized by the expression $W_i = W \cos (2 \pi \alpha i+\phi)$. Here, $W$ stands for the quasiperiodic field strength, $\alpha$ signifies an irrational wavenumber, and $\phi$ represents an arbitrary phase offset.
In the case where $V_i=0$, the Hamiltonian depicted in Eq. (\ref{BHH}) reduces to the non-interacting Aubry-André model~\cite{aubry1980analyticity}, a model which has garnered significant attention due to its localized state properties. Conversely, when $V_i \neq 0$ and with increasing $W$, the system undergoes a notable phase transition from a thermal state to the MBL phase. Within the MBL regime, interactions among particles synergize with the quasiperiodic potential, leading to the localization of the many-body states. This results in intricate quantum dynamics, positioning quasiperiodic potential-driven MBL as a compelling topic in the realm of condensed matter physics~\cite{aubry1980analyticity,PhysRevB.87.134202,PhysRevB.96.075146,PhysRevLett.121.206601}.

\textbf{\em Observation quantities --} To observe the effects of Many Body Localization (MBL) and assess error mitigation in a quantum processor, we consider several physical quantities that distinctly delineate between the MBL phase and the thermalized phase, as supported by various studies. 
%These quantities are invaluable for observing the contrasting behaviors in different phases.

Inverse Participation Ratio (IPR): The IPR quantifies the localization or delocalization degree of a wave function in a many-body system. For a normalized pure state $\psi(t)=\sum c_{n}(t)|n\rangle$, the IPR is defined as \cite{PhysRevLett.115.046603,PhysRevB.83.094431}:
\begin{equation}
IPR=\sum_{n}|c_{n}(t)|^{4},
\end{equation}
IPR, serving as a measure of the Hilbert space distribution of the wave function, yields insights into the system’s localization behavior, aiding in identifying the presence of localized or extended states in complex quantum systems. A higher value of IPR signifies the pronounced localization of the wave function within a confined region of the Hilbert space, attaining its maximal value of 1 when every occupied state is fully localized. In contrast, a diminished IPR value implies a broader dispersal of the wave function across an extensive region of the Hilbert space, suggesting a tendency towards delocalization.

Renyi Entropy: Renyi entropy quantifies the uncertainty or diversity of a system, and it provides insights into the quantum entanglement of the system, with higher values indicating greater entanglement. It is defined as \cite{PhysRevB.77.064426,fan_out--time-order_2017,PhysRevB.92.214204}:
\begin{equation}
S_{n}=\frac{1}{1-n}\log_2 [Tr\rho_{A}^{n}(t)],
\end{equation}
Here, our attention is primarily on the second Renyi entropy $S_{2}=-\log_2 [Tr\rho_{A}^{2}(t)]$. In the ergodic phase, $S(t)$ grows linearly, saturating to the infinite temperature thermal value, whereas, in the MBL phase, it exhibits logarithmic growth, i.e., $S(t)\sim \log(t)$ \cite{PhysRevB.77.064426,PhysRevLett.109.017202,PhysRevLett.115.230401,PhysRevLett.124.243601}.

\begin{figure}[t]
\centering
\includegraphics[width=\columnwidth]{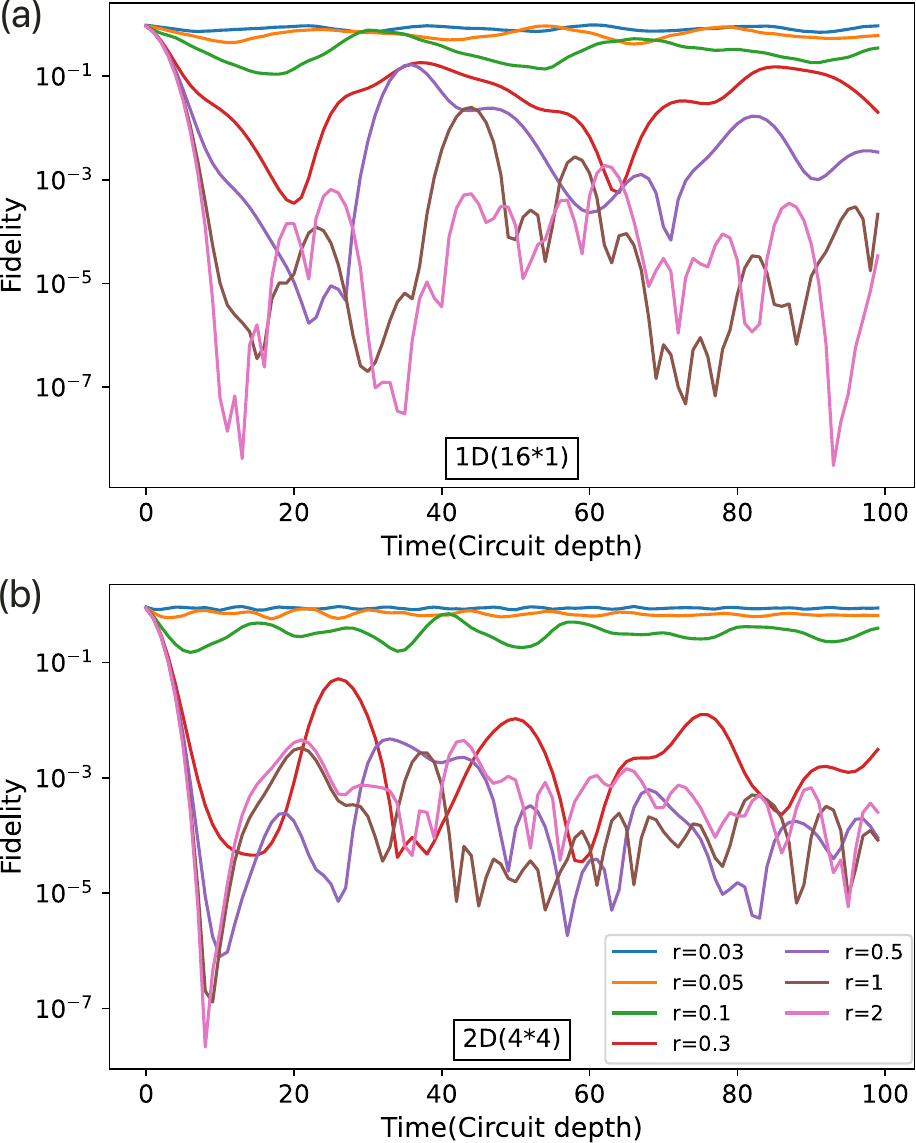}
\caption{Fidelity as a function of time (circuit depth): We explore the time evolution of a multi-qubit quantum processor commencing from a separable state, symbolizing the processor's initial state. Evolution is presented distinctively for idle system scenarios—no gate operations are executed, and all gate controls are idle. The processor configurations evaluated incorporate 16 qubits, organized in both 1D and 2D qubit connectivities (Panel (a) displays 1D, and panel (b) displays 2D). The ratio $r$ illustrates the relationship between the residual interaction's strength and the amplitude of the quasiperiodic potential, portraying transitions from an MBL phase to a thermal phase with its increase. There's a more rapid fidelity decay in the 2D case due to increased neighboring qubit interactions.} \label{fig: evolution}
\end{figure}

\begin{figure}[t]
\centering
\includegraphics[width=\columnwidth]{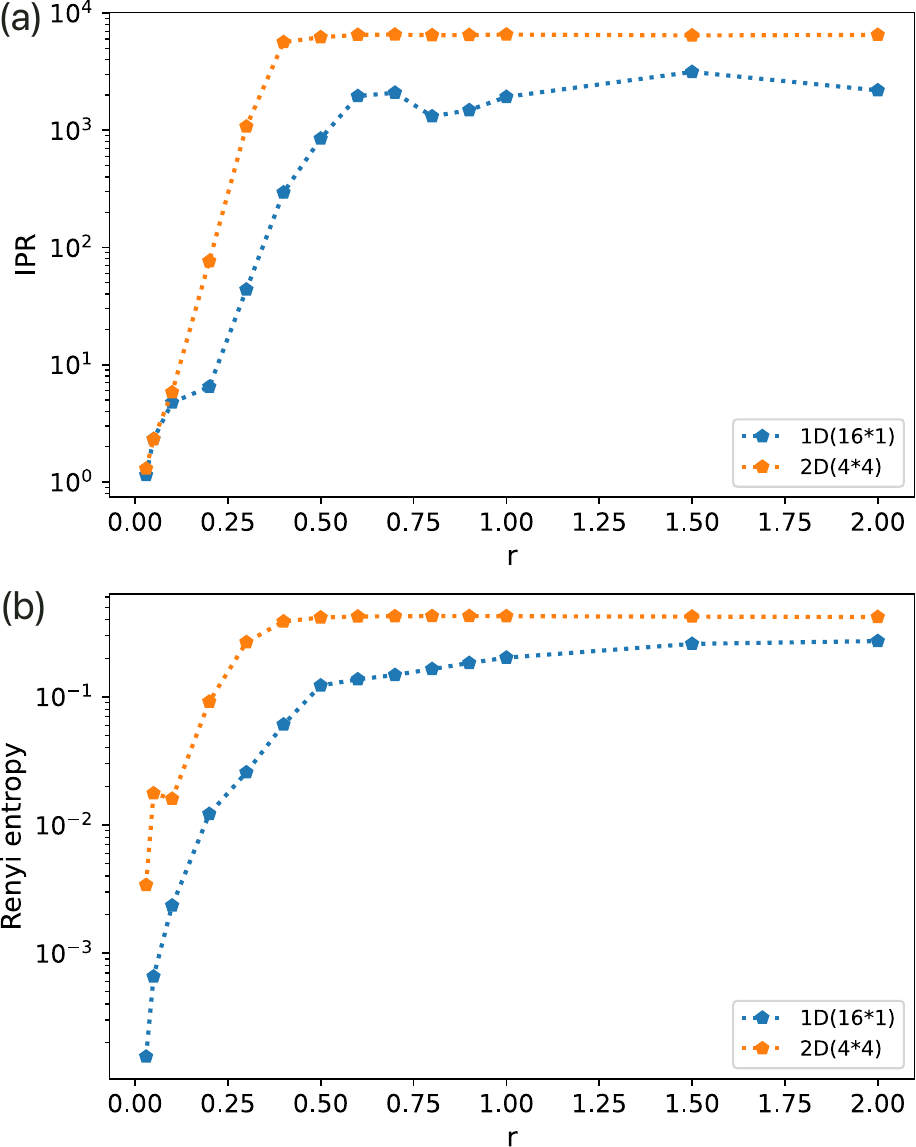}
\caption{IPR is displayed in the upper panel (a), and Renyi entropy is portrayed in the lower panel (b), detailing the contrasts between the terminal and initial states of quantum processors exhibiting both 1D and 2D qubit connectivity. The trends for each qubit-connectivity case are analogous, differentiating the deep-MBL phase from the deep-thermal phase, with IPR/fidelity declining as $r$ amplifies, and Renyi entropy escalating. However, the 2D configuration, with its enhanced neighborhood around a qubit, experiences accelerated alterations.} \label{fig:IPR&RE}
\end{figure}

Fidelity: In the realm of quantum circuits, the evaluation of fidelity is imperative due to the pervasive noise and inevitable control errors within the system. Google utilizes Cross-Entropy Benchmarking (XEB) \cite{boixo2018characterizing,Arute2019Supremacy} as a technique to evaluate the average fidelity of gate operations, comparing the outcomes of actual gate operations against their theoretical counterparts. In our approach, we simulate the circumstances under which XEB circuits operate. Our numerical simulations grant us the ability to circumvent the typical measurement and sampling processes inherent to XEB, enabling the direct computation of fidelity. 

%This is facilitated by our access to the final states of both the simulated and the ideal circuits.

\begin{table*}[ht]
\centering
\caption{Comparison of key quantum computing chips' experimental results with parameter ranges considered in the numerical calculation for MBL and thermal phases in the Bose-Hubbard model.}
\setlength{\tabcolsep}{4.5mm}
\begin{tabular}{p{4cm}<{\centering}|p{10cm}}
\hline \hline
Model & \shortstack[c]{Parameter}
 \\ \hline
\multirow{5}{*}{\centering Google's supremacy~\cite{Arute2019Supremacy}} & \multirow{5}{*}{\centering \shortstack[l]{ residual coupling $h<5$ MHz; \\ average difference between adjacent qubits $|W|\approx 100$ MHz;\\ effective $r<0.1$;\\qubit frequency range $6.5-6.7$GHz}}\\
& \\
& \\
& \\
& \\ \hline
\multirow{5}{*}{\centering IBM~\cite{ding2020systematic}} & \multirow{5}{*}{\centering \shortstack[l]{ residual coupling $h<5$ MHz; \\ average difference between adjacent qubits $|W|\approx500$ MHz; \\effective $r<0.1$\\qubit frequency range $4.9-5.5$GHz}} \\
& \\
& \\
& \\
& \\ \hline
\multirow{5}{*}{\centering Bose-Hubbard model~\cite{PhysRevB.87.134202} }& \multirow{5}{*}{\centering \shortstack[l]{ $h=5$ MHz, $W=h\frac{1}{r}$; \\1D case: $\alpha=\frac{\sqrt{5}-1}{2}$, $\phi=0$; 2D case: $\alpha_{x}=\frac{\sqrt{5}-1}{2}$, $\alpha_{y}=\frac{\sqrt{7}-1}{2}$ \\r<0.5, MBL phase;\\r>0.5, thermal phase }}\\
& \\
& \\ 
& \\
& \\
\hline \hline
\end{tabular}\label{table1}
\end{table*}

\textbf{\em Numerical Results--} We begin by evaluating the efficacy of our mitigation strategies on quantum processors, focusing on the quantities delineated in the preceding section. For such a configuration, the governing model is the Bose-Hubbard model, as outlined previously, please refer to Tab.\ref{table1} for the parameter ranges considered in the numerical calculation and the comparison between those values and those in the real experiments. The implications of IPR and Renyi entropy have been elucidated in Ref.~\cite{PhysRevB.87.134202}, where the authors investigated the long-time dynamics of 1D spin chains of varying lengths. The conclusions drawn establish that, in the presence of a quasiperiodic potential, the system’s phase is determined by the ratio of the hopping interaction strength to the amplitude of the quasiperiodic potential. In the realm of many-body localization, a lower value of the ratio implies that the system is within the MBL phase; however, the existence of a true MBL phase in the thermodynamic limit remains a topic of debate, and our primary focus is predominantly on the prethermal regions of MBL systems in quantum processors with finite system sizes. Conversely, a higher ratio value is indicative of the system being in a thermal phase. It was also inferred that the distinction between the two phases becomes more pronounced as the system size increases.

In our investigation, the system configuration closely mirrors the one mentioned above, with our focus narrowed to the $16$-qubit case. We explore the time evolution of the previously discussed quantities and define the ratio as follows:
\begin{equation}
r=\frac{h}{W}
\end{equation}
Here, $h$ represents the strength of the residual interaction, and $W$ symbolizes the amplitude of the quasiperiodic potential. 

%Our selected initial state is a produce state $|0101,...,0101\rangle$, wherein adjacent sites possess distinct Pauli-$Z$ eigenstates (or energy levels for superconducting qubits). As for the fidelity of the state relative to the initial state, we depict its evolution over $50$ gate times.
%each amounting to $15\text{ns}$. 
%This duration is almost comparable to the qubit lifetime in quantum processors, although it’s crucial to note that our model does not account for noise effects.

%需要说明的是，我们这里qubit的频率是在一个base加上quasiperiodic potential。在1d和2d的情况，我们的base frequency都是$4.889$GHz, 残留耦合$h$的强度都固定为$5$Mhz。$W=\frac{1}{r}h$。对于1d,$\alpha=\frac{\sqrt{5}-1}{2},\phi=0$. 对于2d,$\alpha_{x}=\frac{\sqrt{5}-1}{2},\alpha_{y}=\frac{\sqrt{13}-1}{2}$

\begin{figure*}[t]
\centering
\includegraphics[width=\textwidth]{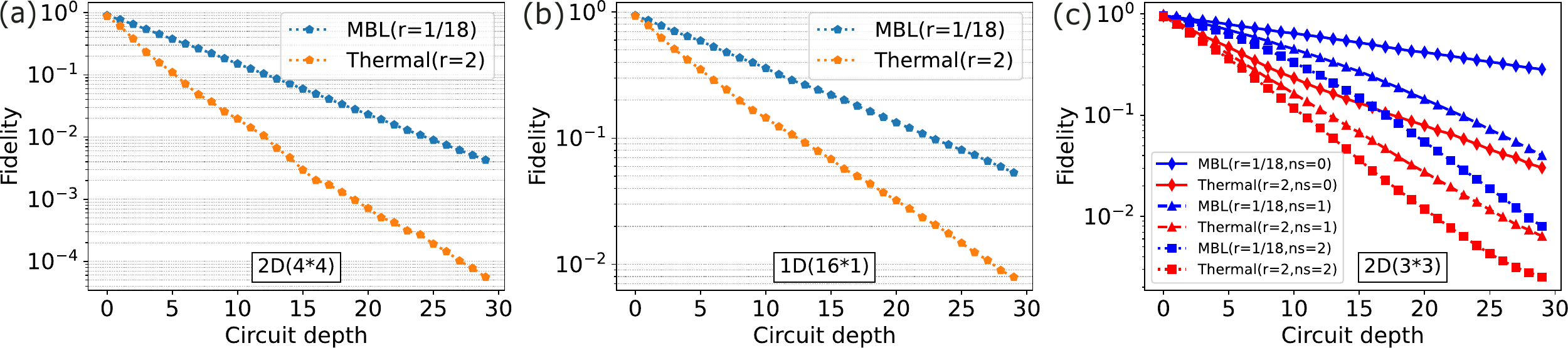}
\caption{This figure depicts an analysis of fidelity within random quantum circuits under varied dimensions and noise conditions, comparing these to their flawless prototypes. Panels (a) and (b) scrutinize scenarios where quantum processors operate in a closed system environment, undisturbed by stochastic noise, encountering only coherent errors due to residual couplings and crosstalks. Both 1D and 2D configurations are evaluated; the latter arranged in a $4 \times 4$ lattice and demonstrating diminished fidelity, attributed to enhanced connectivity between proximate qubits. The color-coded fidelity curves, blue and orange, represent MBL and thermal frequencies respectively, with MBL frequencies markedly elevating fidelity in 2D configurations. Panel (c) incorporates stochastic noise with noise strength (simplified for $ns$) ($ns=0$ represents the pure unitary evolution without stochastic noise; $ns=1$ indicates $T_{1}=50\mu s,T_{2}=69\mu s$; $ns=2$ indicates $T_{1}=25\mu s,T_{2}=34.5\mu s$), specifically $T_1$ and $T_2$ errors, segregating conditions with distinct geometric shapes and colors, reflecting the debilitation of noise mitigation via MBL as noise strength $ns$ increases.
} \label{fig:xeb}
\end{figure*}

In the following numerical calculations, we consider the quantum processors with both 1D and 2D qubit-connectivity with totally $16$ qubits. The initial states in the calculation begin from a separable or product state $|0101,...,0101\rangle$, wherein adjacent sites possess distinct Pauli-$Z$ eigenstates (or different energy levels for superconducting qubits). As for the fidelity of the state relative to the initial state, we depict its evolution over $50$ gate times. Fig.\ref{fig: evolution} and Fig.\ref{fig:IPR&RE} consider in an idle case, meaning no gate operations are executed during the evolution, and all gate controls are idle; while Fig.\ref{fig:xeb} considers the case with random circuits applied as those in XEB experiments.

It is noteworthy that our qubit frequencies undergo modulation with a base frequency and are additionally influenced by a quasiperiodic potential, that is,  $w_{i}=w_{base}+W_{i}$, where $w_{base}$ is base frequency and $W_i = W \cos (2 \pi \alpha i+\phi)$ is the potential value in site $i$. In both the 1D and 2D scenarios, our base frequency remains constant at $w_{base}=4.889$ GHz, while the strength of the residual coupling $h$ is set to $5$ MHz. We tune the parameter $W$ as $W=\frac{1}{r}h$ with $r$ varying. For the 1D case, we adopt $\alpha=\frac{\sqrt{5}-1}{2}$ and $\phi=0$. In the 2D case, we consider $\alpha_{x}=\frac{\sqrt{5}-1}{2}$ and $\alpha_{y}=\frac{\sqrt{7}-1}{2}$.

Here, we emphasize that the achievement of real MBL phase in 2D systems introduces complexities surpassing those in 1D counterparts, primarily attributed to enriched interaction networks and dimensional considerations. However, our primary focus herein pivots towards the pragmatic application of MBL principles for quantum calibration challenges in near-term quantum devices characterized by a finite qubit count and constrained circuit depth. Within such a context, the physics of prethermal phases is perceived to hold more reliable.

In Fig.~\ref{fig: evolution}, the upper panel (a) illustrate the 1D configuration, revealing that in the Many-Body Localized (MBL) phase, where $r < 0.5$, the fidelity oscillates closely around $100\%$. When $r$ is even smaller, fidelity remains notably closer to $1$. In contrast, in thermal-like phases, where $r > 0.5$, fidelity undergoes a rapid decline before experiencing a minor recovery close to $0.1$, and then descending again, indicating an inherent oscillatory behavior but settling at a smaller value compared to the MBL phase. The parameter $r$ plays a important role, depicting the ratio between the residual interaction's strength and the quasiperiodic potential's amplitude, serving as a crucial indicator of the transition from MBL to thermal phases.

In the IPR case, depicted in Fig.~\ref{fig:IPR&RE}(a), similar behaviors are observable for the MBL phase, where $r<0.5$, with the IPR oscillating around $1$. Conversely, in thermal-like phases, where $r>0.5$, the IPR never approaches significant values, indicating signs of thermalization, even when the fidelity obtains substantial values. Deep thermal phase's IPR is nearly four orders of magnitude smaller than deep MBL, affirming substantial deviations from the initial state. Renyi entropy, displayed in Fig.~\ref{fig:IPR&RE}(b), supports these observations, revealing negligible entanglement in the deep MBL phase and significant entanglement in the deep thermal phase.

This detailed representation in Fig.\ref{fig: evolution} and Fig.\ref{fig:IPR&RE} also uncovers  distinctions between 1D and 2D configurations in the manifestation of quantum phenomena.  For the 2D cases, showcased in both Fig.\ref{fig: evolution} and Fig.\ref{fig:IPR&RE}, the accelerated decay in fidelity and increased entanglement can be attributed to the inherent multidirectional interactions and information dissemination facilitated by the 2D lattice structure. The equation
$W_{i} = W \left( \cos (2 \pi \alpha_{x} x_{i}) + \cos (2 \pi \alpha_{y} y_{i}) \right)$
where $x$ and $y$ label the rows and columns respectively, illustrates the relationship between the two different irrational numbers used for quasiperiodic potentials in the 2D configurations.
This 2D architecture, more representative of actual quantum devices, induces faster fidelity decay and heightened entanglement in the thermal phases, as reflected in the observations for different $r$ values in both the 1D and 2D arrangements.
The similarities and differences between the 1D and 2D configurations are coherent across the observed phases, with variations in magnitude and rates due to the dimensional differences. Understanding these subtle dynamics and interactions in quantum processors is pivotal, offering deeper insights into the transitional behaviors between MBL and thermal phases in quantum systems.

Finally, as shown in Fig.\ref{fig:xeb}, with 2D case in $(a)$ and 1D case in $(b)$, we analyze the real-world application of fidelity, applying single and two-qubit gates within a \(20\)-layer XEB random circuit \cite{boixo2018characterizing,Arute2019Supremacy}. Each layer is structured with one single qubit and a two-qubit layer. In the single qubit layer, a random gate from the \( \frac{\pi}{2} \) Pauli gate group \( \{X_{\frac{\pi}{2}}, Y_{\frac{\pi}{2}}, W, I\} \) is applied, where \( W \) represents the square root of \( \frac{(X_{\frac{\pi}{2}}+Y_{\frac{\pi}{2}})}{\sqrt{2}} \). For the two-qubit layer, an ISwap-like gate, symbolizing the evolution of the Hamiltonian, is utilized for four random qubit pairs, considering potential topological constructs.
The depicted blue line, representing a ratio \( r \) of 0.03, resides in the deep MBL phase, maintaining a fidelity order of magnitude of \( 10^{-1} \) even post \(20\) layers. Contrarily, in the deep thermal phase \( r=2 \), fidelity falls below \( 10^{-3} \), establishing a crucial two orders of magnitude difference in the context of practical quantum computation. We further examine the impact of key stochastic noise types, namely $T_1$ and $T_2$ noise models, on Fidelity within both MBL and thermal phase contexts, as illustrated in Fig.\ref{fig:xeb} (c). This analysis underscores the applicability of our schemes in scenarios that more closely mirror realistic conditions.

%It’s observed that gate operations have a pronounced impact on the MBL phase, diverging fidelity behaviors particularly compared to scenarios devoid of gates. This exhaustive exploration illuminates the complexities and interactions within quantum processors, underscoring the substantial influence of dimensions, noise on fidelity, and the intricate role of MBL in counteracting noise, thus providing crucial insights for the progression in quantum computing.

In previous discussions, Fig.~\ref{fig: evolution}  explores fidelity between the quantum chip's final and initial states, sans gate application, leveraging insights from many-body localization (MBL) theories. In the MBL phase, local integrals of motion (LIOMs) emerge, demonstrating a two-body weight nearing unity and suppressing other multi-body terms. 
The Frobenius norm of LIOMs, approaching almost unity, signifies the system's inclination to preserve its initial state. Conversely, the thermal phase sees a rise in non-local multi-body weights of LIOMs, eroding initial information \cite{singh2021local}. 
On the other hand, Fig.~\ref{fig:xeb} scrutinizes the fidelity between the final states of the quantum chip and the ideal circuit under the influence of applied random circuits. Here, fidelity visibly diminishes with increasing circuit depth, unlike situations void of random circuit application. Theoretically, when random circuits are present, the consideration shifts to LIOMs under gate operation influence. MBL's capacity to only preserve the eigenstates of LIOMs~\cite{chandran2015constructing,serbyn2013local,rademaker2016explicit,singh2021local,PhysRevB.82.174411}, and therefore, notably safeguard the state only in idle scenarios between circuit layers, becomes apparent. Moreover, two-qubit gate operations, inducing error propagators between different qubits, might also diminish the fidelity of quantum processor. Yet, when compared with thermal phase scenarios, MBL's role in boosting circuit fidelity is apparent, affirming our scheme's efficacy in notably mitigating effects from residual coupling errors.

\textbf{\em Conclusions and discussions--}
This research presents a approach for mitigating crosstalk and residual coupling errors in superconducting quantum processors, focusing on leveraging the principles of Many-Body Localization (MBL) to enhance the precision and efficiency of qubit frequency control. Superconducting quantum processors belong to a key platform for the advancement of quantum computers, but their optimum performance is intertwined with meticulous calibration and parameter selection, primarily concerning qubit frequency control.
Our proposed MBL-based methodology demonstrates substantial resilience to noise and residual coupling, optimizing performance and addressing the inherent challenges of superconducting chip calibration, such as gate control errors and instabilities. Unlike existing strategies, our approach facilitates convenience and cost savings in the calibration process, reducing the requirement for extensive optimization computation, even as the quantum system scales.
We believe this study elucidate strategies for enhancing the fidelity, stability, and efficacy of superconducting qubits. In addition, our result is poised to advance calibration methodologies and optimize the performance of superconducting qubits, offering a robust framework and insights for forthcoming investigations into calibration and error mitigation within superconducting quantum processors.

In conclusion, it is imperative to underscore certain potential limitations inherent to our MBL-based mitigation scheme. Initially, the existence of a stable MBL phase, particularly in the contexts of extended durations and considerable system sizes, remains a subject of ongoing debate. Despite this, it is crucial to note that our methodology could still serve as a robust scheme for near-term quantum devices characterized by finite qubit numbers and finite circuit depth. In such systems, the prethermal phase in those disordered interacting systems is conjectured to possess considerable robustness. Subsequently, the presence of pronounced stochastic noise could potentially compromise the localization phase, necessitating a conscientious restriction of the strength of the stochastic noise, specifically, $T_1$ and $T_2$ noise, to a minimal. Within the confines of a weak noise regime, our approach has demonstrated its validity and efficiency as shown in Fig.\ref{fig:xeb}(c).

In conjunction with these important considerations, a deeper insight into the intricate dynamics and interactions in quantum processors across varied configurations will facilitate the development of more refined strategies, providing robust solutions to prevalent challenges in quantum computing. Moreover, in light of ongoing advancements in quantum technologies, integrating novel technologies and methodologies stands to augment the efficacy of our proposed approach, enabling its application in a wider array of quantum computing architectures with increased comprehensiveness.

%Finally, we want to emphasize some potential limitations of our MBL-based mitigation scheme. First of all, the existence of stable MBL phase in the long-time and large size limit is still under debate. While we emphasize that our method could still provide a valuable strategy for near-term quantum devices with finite qubit number and finite circuit depth, where the pretherml phase in those disorder interacting systems are believed to be very robust. Secondly, the localization phase could be destroyed by strong stochastic noise, therefore we have to limit the strength of the stochastic nosie, i.e. $T_1$ and $T_2$ noise, to be weak. In the weak noise regime, we did show that our scheme are valid and efficient. Along with those crucial point, we believe that the exploration of additional modifications and improvements to enhance the resilience and adaptability of our approach is crucial. Diving deeper into understanding the intricate dynamics and interactions in quantum processors under varied configurations will lead to refined strategies, offering more robust solutions to the challenges faced in the realm of quantum computing. Moreover, considering the advancements in quantum technologies, the integration of emerging technologies and methodologies could potentially amplify the capabilities of our proposed approach, allowing for more comprehensive and versatile applications in diverse quantum computing architectures.

\begin{acknowledgements}
\textbf{\em Acknowledgments.}---D.E.L. is supported by the National Natural Science Foundation of China (Grant No.~92365111), the Beijing Natural Science Foundation (No.~Z220002), and the Innovation Program for Quantum Science and Technology (Grant No.~2021ZD0302400).
X.L. is supported by the Research Grants Council of Hong Kong (Grants No.~CityU~21304720, No.~CityU~11300421, No.~CityU~11304823, and No.~C7012-21G) and City University of Hong Kong (Projects No.~9610428 and No.~7005938).
\end{acknowledgements}

\bibliography{ref}
\end{document}